\documentclass[10pt,conference]{IEEEtran}
\pdfoutput=1
\IEEEoverridecommandlockouts

\usepackage[nocompress]{cite}
\usepackage{algorithm}
\usepackage[noend]{algpseudocode}

\usepackage[caption=false,labelfont=sf,textfont=sf]{subfig}
\usepackage[cmex10]{amsmath}
\usepackage[pdftex]{graphicx}
\usepackage{url}
\usepackage{amsfonts}
\usepackage{flushend}
\usepackage[hyperfootnotes=false]{hyperref}
\hypersetup{
colorlinks=true,
citecolor=blue,
linkcolor=blue,
urlcolor=blue
}

\begin{document}

\title{CARP: Context-Aware Reliability Prediction\\ of Black-Box Web Services\vspace{1ex}}

\author{\IEEEauthorblockN{Jieming Zhu\textsuperscript{\scriptsize 1},
~~Pinjia He\textsuperscript{\scriptsize 1},
~~Qi Xie\textsuperscript{\scriptsize 2}\IEEEauthorrefmark{1}\thanks{\IEEEauthorrefmark{1}\textit{The work was done when the author was visiting CUHK.}},
~~Zibin Zheng\textsuperscript{\scriptsize 3},
~~Michael R. Lyu\textsuperscript{\scriptsize 1}
\vspace{1ex}
}
\IEEEauthorblockA{\textsuperscript{\scriptsize 1}Shenzhen Research Institute, The Chinese University of Hong Kong, China\\
Department of Computer Science and Engineering, The Chinese University of Hong Kong, Hong Kong\\
\textit{\{jmzhu, pjhe, lyu\}@cse.cuhk.edu.hk}\vspace{1ex}\\
\textsuperscript{\scriptsize 2}School of Computer Science and Technology, Southwest Minzu University, China\\
\textit{qi.xie.swun@gmail.com}\vspace{1ex}\\
\textsuperscript{\scriptsize 3}Key Laboratory of Machine Intelligence and Advanced Computing (Sun Yat-sen University), Ministry of 
Education\\School of Data and Computer Science, Sun Yat-Sen University, China\\
\textit{zhzibin@mail.sysu.edu.cn}}
}

\maketitle

\begin{abstract}

Reliability prediction is an important task in software
reliability engineering, which has been widely studied in the
last decades. However, modelling and predicting user-perceived
reliability of black-box services remain an open research problem.
Software services, such as Web services and Web APIs, generally
provide black-box functionalities to users through the Internet,
thus leading to a lack of their internal information for reliability
analysis. Furthermore, the user-perceived service reliability
depends not only on the service itself, but also heavily on the
invocation context (e.g., service workloads, network conditions),
whereby traditional reliability models become ineffective and
inappropriate. To address these new challenges posed by blackbox
services, in this paper, we propose CARP, a new contextaware
reliability prediction approach, which leverages historical
usage data from users to construct context-aware reliability
models and further provides online reliability prediction results
to users. Through context-aware reliability modelling, CARP is
able to alleviate the data sparsity problem that heavily limits the
prediction accuracy of other existing approaches. The preliminary
evaluation results show that CARP can make a significant
improvement in reliability prediction accuracy, e.g., about 41\% in
MAE and 38\% in RMSE when only 5\% of the data are available.
\end{abstract}


\section{Introduction}\label{sec:intro}

Reliability measures the probability of failure-free software operation for a specified period of time in a specified environment~\cite{Lyu96}. Reliability prediction is an important task in software reliability engineering~\cite{Lyu96, Lyu07}, which aims to predict failure rates of components and overall system reliability. These predictions are commonly used to evaluate design decisions, trade-off design factors, identify potential failure areas, and track reliability improvement~\cite{ReliabilityPredictionBasics}. In the last few decades, reliability prediction has been widely studied, producing a variety of prediction models (e.g., Palladio component model~\cite{BroschKBR12} and Poisson process model~\cite{HuangLK03}). However, most of these existing models target at reliability analysis of traditional white-box software systems, where the reliability of system components are known or can be estimated through behaviour models from internal information of the components. How to model and predict the user-perceived reliability of emerging Web services remains an open reserach problem.

Nowadays, various software services such as Web services and Web APIs are emerging over the Internet. These services have become an integral part for building modern Web applications, in which each service provides a black-box functionality via some standard interfaces. To evaluate the reliability of a (third-party) black-box service, traditional white-box reliability prediction approaches become inapplicable due to a lack of its internal behaviour information. In addition, different from stand-alone software systems, software services operate over the Internet and likely serve different users spanning worldwide~\cite{ZhuZZL13_Scale}. Therefore, the user-perceived reliability may differ from user to user due to different user locations, and vary from time to time due to dynamic service workloads and network conditions. In such a setting, it is more suitable to evaluate service reliability from user side than from system side as evaluating traditional software systems. As a result, modelling and predicting user-perceived reliability of black-box services is an important task, which is exactly the goal of our work.

Specifically, as with~\cite{SilicDS13, ZhengL10}, we compute user-perceived service reliability as the ratio of the number of successful service invocations against the total number of service invocations performed by the user. The most straightforward way, therefore, is to assess the reliability of a target service through real invocations from users. However, each service usually has many users and each user may need to assess a lot of alternative services (with similar or identical functionalities). Such exhaustive invocations can impose additional cost for users (e.g., the service invocations may be charged) and also incur expensive overhead for service systems (e.g., by consuming additional system resources), thus making it infeasible in practice. It is more desirable to identify approaches that can achieve accurate reliability predictions without requiring additional service invocations.

Towards this end, a few initial efforts have been made by several recent studies, by applying the K-means clustering technique~\cite{SilicDS13,SilicDS15} or collaborative filtering techniuqes used in recommender systems~\cite{ZhengL10, ZhengL13}. These studies collect partial invocation data (i.e., observed reliability on the invoked services) from users and then construct statistical models for prediction of unknown reliability records. Whereas these approaches obtain encouraging results, two significant challenges remain: \textit{1) Context modelling.} From a user's perspective, reliability not only depends on a service itself, but also is highly influenced by the context of service invocations (e.g., service load and network conditions). For example, user-perceived reliability may differ from user to user due to different user locations, and may vary from time to time incurred by service load variations and dynamic network conditions. How to leverage such context information to aid in reliability prediction is still a challenging problem. \textit{2) Data sparsity.} In practice, each user typically invokes only a few services at each time, leading to a limited number of invocation samples. When modelling the reliability given a specific context, the data matrix becomes extremely sparse (i.e., most of reliability records are unknown). With limited training data, it is difficult to make accurate reliability predictions. 

In this paper, we propose {CARP}, a \underline{c}ontext-\underline{a}ware \underline{r}eliability \underline{p}rediction approach that aims to tackle the above challenges. {CARP} models reliability as a function jointly determined by the (\textit{user, service, context}) tuple of a service invocation. Then the model is constructed based on a novel formulation of context-specific matrix factorization by lerveraging the implicit context information between users and services. To guarantee computational efficiency, {CARP} comprises an \textit{offline} step to train the context-aware reliability model from historical invocation data, and another \textit{online} step to support on-demand reliability predictions for ongoing service invocations. We evaluate {CARP} on a publicly available dataset with real-world reliabiltiy samples collected from Amazon EC2 services~\cite{SilicDS13}. The evaluation results show that {CARP} can better capture the characteristics inherent in  reliability of black-box Web services and therefore yields a significant improvement in prediction accuracy (e.g., up to 41\% in MAE and 38\% in RMSE) over the state-of-the-art reliability prediction models. 

The main contributions of this paper can be summarized as follows:
\begin{itemize}\setlength\itemsep{1ex}
\item We study the problem of predicting user-perceived reliability of black-box Web services, which remains an open and challenging research problem.
\item We present {CARP}, a context-aware reliability model with its construction for reliability prediction, by leveraging a novel formulation of context-specific matrix factorization. 
\item The evaluation results show that {CARP} makes a significant improvement in prediction accuracy over the state-of-the-art reliability prediction models.
\end{itemize}

The remainder of this paper is organized as follows. Section~\ref{sec:background} introduces the background. Section~\ref{sec:approach} describes the details of {CARP}. We report the evaluation results in Section~\ref{sec:evaluation} and make some discussions in Section~\ref{sec:discussion}. We then review the related work in Section~\ref{sec:relatedwork}, and finally conclude the paper in Section~\ref{sec:conclusion}.

\begin{figure*}[!t]
  \centering
  \includegraphics[width=1 \textwidth]{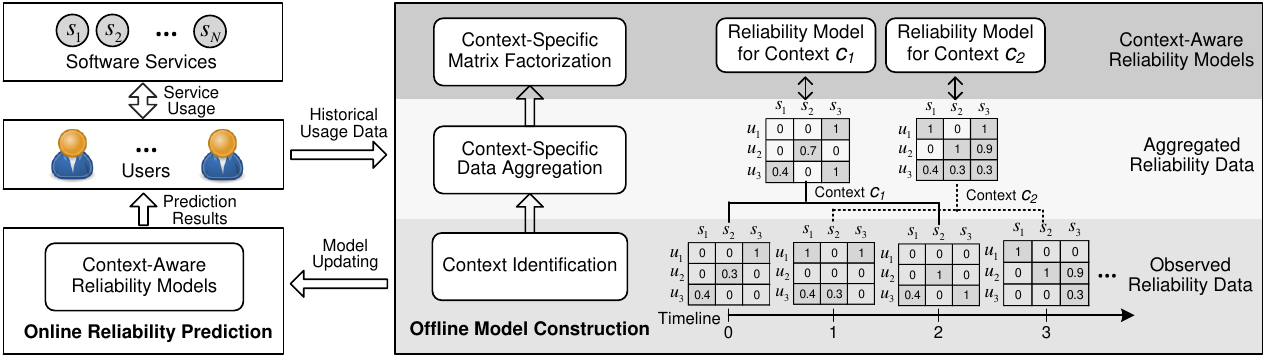}
  \caption{The Framework of Context-Aware Reliability Prediction}\label{fig:framework}
\end{figure*}

\section{Background}\label{sec:background}
Collaborative filtering (CF) techniques~\cite{SuK09} are widely used to rating prediction in recommender systems, such as movie recommendation in Netflix.  The goal of CF is to leverage partially-observed rating data to predict the remaining unknown ratings, so that movies can be recommended to users according to the predicted ratings. Matrix factorization (e.g., PMF~\cite{SalakhutdinovM07}) is a classic model to address the collaborative filtering problem, which constrains the rank of the data matrix, \textit{i.e.}, $rank(R) = d$. The low-rank assumption is based on the fact that the entries of $R$ are largely correlated, thereby resulting in a low effective rank in $R$. Concretely, factoring a matrix is to map both users and services into a joint latent factor space of a low dimensionality $d$ such that values of the data matrix can be captured as inner products of latent factors in that space. Then the latent factors can be employed for further prediction on unknown data entries. 

Formally, given $n$ users and $m$ services, we denote latent user factors as $U \in \mathbb{R}^{d \times n}$ and latent service factors as $S \in \mathbb{R}^{d \times m}$. Both of them are used to fit the data matrix $R$, \textit{i.e.}, $R \approx U^TS$. To avoid overfitting, regularization terms that penalize the norms of the solutions (\textit{i.e.}, $U$ and $S$) are added. Thus we resolve to minimize the following loss function:
\begin{equation}\label{equ:MF}
\small
\mathcal{L} = \frac{1}{2}\sum\limits_{i = 1}^n \sum\limits_{j = 1}^m I_{ij}{{(R_{ij} - U^T_i{S_j})}^2} + \frac{\lambda_U}{2}\left\| U \right\|_F^2 + \frac{\lambda_S}{2}\left\| S \right\|_F^2,
\end{equation}
where the first term indicates the sum of squared error in approximation. Especially, $I_{ij}$ acts as an indicator that equals to 1 if $R_{ij}$ is observed, and 0 otherwise. The remaining terms, namely regularization terms, are added to avoid overfitting. ${\left\|  \cdot  \right\|_F}$ denotes the \textit{Frobenius norm}~\cite{SalakhutdinovM07}, and $\lambda_U, \lambda_S$ are two parameters to control the extent of regularization.

Gradient descent is a widely used method to find a local minimum of an objective function in an iterative way. As for the PMF model expressed in Equation~\ref{equ:MF}, the gradient descent algorithm works by initializing $U_i$ and $S_j$ randomly and iterating over the following updating rules until convergence:
{\small
\begin{eqnarray}
{U_i} \leftarrow {U_i} - \eta \frac{{\partial \mathcal{L}}}{{\partial {U_i}}},\\
{S_j} \leftarrow {S_j} - \eta \frac{{\partial \mathcal{L}}}{{\partial {S_j}}},
\end{eqnarray}}
\hspace{-0.7ex}After obtaining the derivatives of $U_i$ and $S_j$ from Equation~\ref{equ:MF}, we derive the following updating rules:
{\small
\begin{eqnarray}
{U_i} \leftarrow {U_i} - \eta \big (\sum\limits_{j = 1}^m {{I_{ij}}(U_i^T{S_j} - {R_{ij}}){S_j} + {\lambda _U}{U_i}} \big ),\\
{S_j} \leftarrow {S_j} - \eta \big (\sum\limits_{i = 1}^n {{I_{ij}}(U_i^T{S_j} - {R_{ij}}){U_i} + {\lambda _S}{S_j}}\big ).
\end{eqnarray}}
\hspace{-0.7ex}In this way, the latent factors $U_i$ and $S_j$ move iteratively by a small step of the average gradients, \textit{i.e.}, $\frac{{\partial \mathcal{L}}}{{\partial {U_i}}}$ and $\frac{{\partial \mathcal{L}}}{{\partial {S_j}}}$, where the step size is controlled by a learning rate $\eta$.

\section{Context-Aware Reliability Prediction}\label{sec:approach}
In this section, we describe the overview and details of our context-aware reliability prediction approach.

\subsection{Overview}
Fig.~\ref{fig:framework} presents our context-aware reliability prediction framework, which comprises three phases: \textit{1) Data collection}. A user-collaboration mechanism, proposed in our previous work~\cite{ZhengL08}, is applied to collecting historical usage data from users. Users can contribute their observed reliability data on the invoked services and get back personalized (i.e., from user side) reliability prediction results. \textit{2) Offline model construction}. Using the collected reliability data, we can construct the context-aware reliability model by a process involving context identification, context-specific data aggregation, and context-specific matrix factorization. The model construction can be performed offline at a periodical interval to update the model parameters with newly-observed reliability data. \textit{3) Online reliability prediction}. The constructed reliability models can be used to provide personalized reliability prediction results to users in an online manner.

\subsection{Context-Aware Reliability Model}
For traditional software systems, the researchers generally take reliability as a constant value that measures the probability of failure-free software operation. Given a specified period of time, the software reliability is defined as follows:
\begin{equation}
\small
r(s),
\end{equation}
where $s$ denotes the specific software system or component. $r(s)$ depends on the software-specific parameters such as software architecture, system resources (e.g., CPU, memory, and I/O), and other software design and implementation factors.

However, this traditional reliability model is not applicable for measuring user-perceived reliability of black-box services. As mentioned before, service reliability should be evaluated from user side other than from system side as evaluating traditional software systems. Due to the influence of user locations and network connections, different users may experience quite different reliability even on the same service. To characterize user-perceived reliability, Zheng et al.~\cite{ZhengL08, ZhengL10} propose the following model:
\begin{equation}
\small
r(u, s),
\end{equation}
where $u$ and $s$ denote the specific user and service respectively. $r(u, s)$ depends both on user $u$ and service $s$.

Further, this model is extended to incorporate temporal information in~\cite{SilicDS13}, considering that user-perceived reliability may vary from time to time due to fluctuating service workloads and dynamic network conditions. Specifically,
\begin{equation}
\small
r(u, s, t),
\end{equation}
formulates the user-perceived reliability for an invocation $inv(u, s, t)$ between user $u$ and service $s$ at time slice $t$.

Although this new model, $r(u, s, t)$, can naturally characterize user-perceived reliability well, we find it difficult to directly apply it to reliability prediction. Because each user has limited historical usage data, applying $r(u, s, t)$ to model the data can lead to such data sparsity problem and thus result in inaccurate predictions. 

In this paper, we argue that the time-dimensional characteristics can be typically captured by a finite set of context conditions, each of which is an abstract representation of the underlying factors such as service workloads and network conditions. It is further endorsed by the fact that service workloads and network conditions likely have regular daily distributions~\cite{WangLS09}. Thus, a specific context condition likely determines the reliability value at a specific time slice. Based on this observation, we propose a context-aware reliability model:
\begin{equation}
\small
r(u, s, c),
\end{equation}
where $c$ denotes the specific context condition under which the invocations $inv(u, s, t)$ are performed. $r(u, s, c)$ indicates that the user-perceived reliability depends on the user $u$, the service $s$, and the context $c$. Especially, $r(u, s, t) \approx r(u, s, c)$, if the context condition is $c$ at time slice $t$. With a limited number of contexts, this model can make reliability data less sparse. In the following, we will describe the use of this model for context-aware reliability prediction.

\subsection{Offline Model Construction}
Formally, we can collect a 3-dimensional matrix $R \in \mathbb{R}^{M \times N \times T}$, which records the reliability data for $M$ users, $N$ services, and $T$ time slices. $R_{u,s,t}$ = $r(u, s, t)$ when the reliability value $r(u, s, t)$ of invocations $inv(u, s, t)$ is observed; otherwise, we set $R_{u,s,t}$ = $0$ as an unknown entry. Due to the afore-mentioned data sparsity problem, the matrix $R$ is highly sparse in practice with a large number of unknown entries. The goal of reliability prediction is to predict these unknown entries, whereby the reliability of an ongoing invocation can be further predicted. As illustrated in the right panel in Fig.~\ref{fig:framework}, the offline  model construction comprises three steps: \textit{context identification}, \textit{context-specific data aggregation}, and \textit{context-specific matrix factorization}. 

\vspace{1ex}
\textit{\textbf{1) Context Identification}}

To characterize and identify different context conditions, we employ \textit{k-means} clustering technique to cluster the reliability data $R$ with $T$ time slices into $C$ clusters, where each cluster represents a specific context and different time slices grouped into one cluster belong to the same context. To achieve this, the observed reliability data between $M$ users and $N$ services at each time slice $t$ can be constructed as a feature vector for \textit{k-means} clustering. However, due to the sparse nature of $R$, the feature vectors would become high-dimensional and sparse, further leading to bad clustering performance. To overcome this issue, we define a feature vector $f_t$ for time slice $t$ using the average reliability value of each service:
\begin{equation}
\small
f_t = \big(\bar{r}(s_1, t), \bar{r}(s_2, t), \cdots, \bar{r}(s_N, t)\big),
\end{equation}
where $\bar{r}(s, t) = mean(\{R_{u,s,t}~|~R_{u,s,t} > 0,~1 \leq u \leq M\})$ calculates the average reliability value of service $s$ over the observed entries at time slice $t$. Using these feature vectors, we perform data clustering and get $C$ different context conditions.

\vspace{1ex}
\textit{\textbf{2) Context-Specific Data Aggregation}}

Different time slices may be clustered into each context. To alleviate the data sparsity problem, we propose to aggregate the data of different time slices within the same context. An aggregated data matrix $\bar{R} \in \mathbb{R}^{M \times N \times C}$ can thus be obtained, where each entry $\bar{R}_{u, s, c}$ denotes the average reliability value between user $u$ and service $s$ in context $c$:
\begin{equation}
\small
\bar{R}_{u, s, c} = mean(\{R_{u, s, t}~|~R_{u, s, t}>0,~t \in context~c\})
\end{equation}
Especially, $\bar{R}_{u, s, c} = 0$ indicates that the reliability for invocations $inv(u, s, t)$ performed in context $c$ is unknown. For example, in Fig.~\ref{fig:framework}, the observed reliability data of four time slices are aggregated into two contexts (i.e., context $c_1$ and $c_2$) and thus the aggregated data become much denser. 

\vspace{1ex}
\textit{\textbf{3) Context-Specific Matrix Factorization}}

The problem of context-aware reliability prediction is to predict the unknown entries (where $\bar{R}_{u, s, c} = 0$) of the aggregated data $\bar{R}$. This can be modelled as a collaborative filtering (CF) problem, which aims for recovering the full matrix from a small number of observed entries. Taking Fig.~\ref{fig:framework} as an example, in the aggregated matrix for context $c_1$, we have four entries observed (e.g., $\bar{R}_{u_3, s_1, c_1} = 0.4$) and five unknown entries to predict (e.g., $\bar{R}_{u_1, s_1, c_1}$). Matrix factorization (MF)~\cite{SalakhutdinovM07} is a classic CF model that allows for low-rank matrix approximation. Different with conventional matrix factorization, we have a 3-dimensional reliability data matrix $\bar{R}\in \mathbb{R}^{M \times N \times C}$, including one 2-dimensional $M$-by-$N$ data matrix $\bar{R}^{(c)}$ in each context $c$ ($1 \leq c \leq C$), where its entry $\bar{R}^{(c)}_{u,s} = \bar{R}_{u, s, c}$.

In such a setting, we propose context-specific matrix factorization. Formally, factorizing a data matrix $\bar{R}^{(c)} \in \mathbb{R}^{M \times N}$ is to map both users and services into a $d$-dimensional latent factor space, such that the values of $\bar{R}^{(c)}$ can be captured as the inner products of the corresponding latent factors $U^{(c)} \in \mathbb{R}^{d \times M}$ and $S^{(c)} \in \mathbb{R}^{d \times N}$, i.e., $\bar{R}^{(c)} \approx {U^{(c)}}^TS^{(c)}$, where ${U^{(c)}}^T$ is the transpose of $U^{(c)}$. Therefore, the context-specific MF model for context $c$ is to minimize the following loss function:
\begin{equation}\label{equ:mf}
\small
\hspace{-0.1ex}\mathcal{L}^{(c)} \hspace{-0.5ex}= \hspace{-0.5ex}\frac{1}{2}\sum\limits_{u, s} I^{(c)}_{us}{{\big (\bar{R}^{(c)}_{us} - {U^{(c)}_u}^T{S^{(c)}_s}\big )}^2} \hspace{-0.9ex}+ \hspace{-0.1ex}\frac{\lambda}{2} \big (\left\| U^{(c)} \right\|_F^2 \hspace{-0.9ex}+\hspace{-0.1ex} \left\| S^{(c)} \right\|_F^2 \big ),\hspace{-1ex}
\end{equation}
where the first term measures the sum of the squared errors between the observed value $\bar{R}^{(c)}_{us}$ and the estimated value ${U^{(c)}_u}^T{S^{(c)}_s}$, and the second is a regularization term used to avoid the overfitting problem~\cite{SalakhutdinovM07}. $I^{(c)}_{us}$ acts as an indicator: $I^{(c)}_{us} = 1$ if $\bar{R}^{(c)}_{us}>0$; $I^{(c)}_{us} = 0$, otherwise. $\lambda$ is a parameter to control the extent of regularization.

The algorithm of \textit{gradient descent}~\cite{SalakhutdinovM07} is usually employed to solve the MF model in Equation~\ref{equ:mf}. For ease of computation, we solve each context-specific MF model sequentially, and employ the solution of the last context for initialization of the current one (e.g., use $U^{(1)}$ and $S^{(1)}$ to initialize $U^{(2)}$ and $S^{(2)}$). At last, we can obtain a pair of $U^{(c)}$ and $S^{(c)}$ for each context $c$. In practice, the offline model construction can be performed periodically to update the models with newly-observed reliability data.

\subsection{Online Reliability Prediction}
The constructed models (i.e., $U^{(c)}$ and $S^{(c)}$) allow for reliability prediction for invocations performed between user $u$ and service $s$ in context $c$, i.e., $\hat{R}_{u, s, c} = {U^{(c)}_u}^TS^{(c)}_s$, where $\hat R$ denotes the predicted matrix corresponding to $\bar R$. This is the basis for performing online reliability prediction, which aims to predict the user-perceived reliability of an ongoing invocation $inv(u, s, t_c)$. Therefore, we seek to associate the invocation context at the current time slice $t_c$ to an existing context $c$. In our implementation, we use the newly observed reliability data to help identify the current context. Specifically, given the observed feature vector $f_{t_c} = \big(\bar{r}(s_1, t_c), \bar{r}(s_2, t_c), \cdots, \bar{r}(s_N, t_c)\big)$, we group it into one of the existing context clusters. After obtaining the context $c$, the reliability of $inv(u, s, t_c)$, denoted as $\hat{r}(u, s, t_c)$, can be predicted by $\hat{r}(u, s, t_c) = \hat{R}_{u, s, c}$.

\section{Evaluation}\label{sec:evaluation}
In this section, we present our results on evaluating the effectiveness of \textit{CARP}. For ease of reproducing our approach, we release our source code with detailed experimental results on our WS-DREAM project page\footnote{{\url{http://wsdream.github.io/CARP}}}. The WS-DREAM repository~\cite{wsdream_repo} is currently hosted on Github to disseminate our research results as well as to release open datasets and source code for Web service research. With both datasets and source code publicly released, our WS-DREAM repository would allow easily reproducing the existing approaches and give flexibility of extending new ones, which hopefully inspires more research efforts in the Service Society.

\subsection{Data Description}
\begin{figure}[!t]
\begin{minipage}[b]{0.5\linewidth}
\footnotesize\centering
\begin{tabular}{l|l}
\hline
\textbf{Statistics}&\textbf{Values}\\
\hline 
\hline \#Records &17,150\\
\hline \#Users &50\\
\hline \#Services & 49\\
\hline \#Workloads & 7\\
\hline Reliability range & $0\sim1$\\
\hline Reliability average & $0.433$\\
\hline
\hline
\end{tabular}
\vspace{0.1in}
\caption{Data Statistics}\label{tab:dataset} 
\end{minipage}
\begin{minipage}[b]{0.47\linewidth}
\centering
\includegraphics[width=1\textwidth]{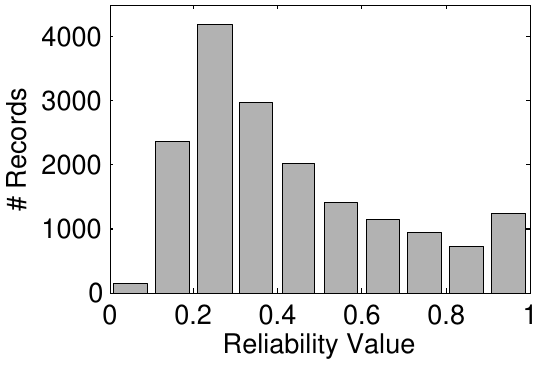}
\vspace{-0.25in}
\caption{Data Distribution}
\label{fig:datadistribution}
\end{minipage}
\end{figure}

Our experiments are conducted based on a real-world reliability dataset recently released in~\cite{SilicDS13}. The dataset was collected using Amazon EC2 platform, which contains 17,150 reliability records from about 2.5 million invocations between 50 users and 49 services under 7 different workloads. Fig.~\ref{tab:dataset} and~\ref{fig:datadistribution} present some data statistics and the data distribution. Specifically, the services are implemented as matrix multiplication operations with different computational complexities, while the users are simulated by a ``stress testing" tool, \textit{loadUI}~\cite{SilicDS13}. Both users and services are deployed into different locations across the seven EC2 regions. The service workload is controlled by setting different time intervals (i.e., 3$\sim$9 sec) between consecutive invocations. Each reliability value is calculated as the successful ratio of 150 consecutive service invocations.

\begin{table}[!t]
\centering \caption{Accuracy Comparison} \label{tab:comparison}
\vspace{0ex}
\begin{tabular}{c|c||c|c|c|c|c}
\hline
&&\multicolumn{5}{c}{Data Density}\\
\cline{3-7}
\hspace{-1ex}\raisebox{1.5ex}[0pt]{Metric}\hspace{-1ex} & \raisebox{1.5ex}[0pt]{Approach} &   5\%    & 10\%  &   15\%   & 20\% &  25\% \\
\hline
\hline
&   Baseline   &   0.176  &   0.170  &  0.168  &  0.169  &  0.170   \\
&   Hybrid~\cite{ZhengL10}  &   0.152  &  0.084  &  0.079  &  0.073  &  0.073   \\
&   CLUS~\cite{SilicDS13}  &   0.077  &   0.059  &  0.043  &  0.036  &  0.031    \\
&   PMF~\cite{ZhengL13}  &  0.076  &  0.031  &  0.021  &  0.017  &  0.014 \\
&   \textbf{CARP}  &  \textbf{0.045}  &   \textbf{0.022}  &   \textbf{0.017}  &   \textbf{0.014}  &   \textbf{0.013}   \\
\cline{2-7}
\hspace{-1ex}\raisebox{6ex}[0pt]{MAE}\hspace{-0.5ex} & Impr.(\%) &  41.0\%  &  27.8\%  &   20.3\%  &   13.9\%  &   13.2\% \\
\hline
\hline
&   Baseline   &   0.217  &   0.211  &  0.210  &  0.210  &  0.211   \\
&   Hybrid~\cite{ZhengL10}  &   0.204  &  0.109  &  0.102  &  0.094  &  0.094  \\
&   CLUS~\cite{SilicDS13}  &  0.112  &    0.093   &   0.066   &   0.060     &  0.052    \\
&   PMF~\cite{ZhengL13}  &   0.110  &   0.050  &    0.036  &    0.031  &    0.028   \\
&   \textbf{CARP} &  \textbf{0.067}  &  \textbf{0.037}  &   \textbf{0.031}  &   \textbf{0.029}  &   \textbf{0.027}   \\
\cline{2-7}
\hspace{-1ex}\raisebox{6ex}[0pt]{RMSE}\hspace{-0.5ex} & Impr.(\%) &  38.9\%  &  24.9\%  &   14.8\%  &   5.8\%  &   6.0\% \\
\hline
\hline
\end{tabular}
\end{table}
\subsection{Evaluation Metrics}
To evaluate prediction accuracy, we use two standard error metrics, MAE (Mean Absolute Error) and RMSE (Root Mean Square Error):
\begin{equation}
\small
MAE = {{\sum\limits_{inv(u, s, t)} {\big| {\hat{r}(u, s, t) - {r(u, s, t)}} \big|} } \big{/}N},
\end{equation}
\begin{equation}
\small
RMSE = \sqrt{{{\sum\limits_{inv(u, s, t)} {\big( {\hat{r}(u, s, t) - {r(u, s, t)}} \big)^2} } \big{/}N}},
\end{equation} 
where $r(u, s, t)$ and $\hat{r}(u, s, t)$ denote the observed reliability value and the corresponding predicted value respectively, for an invocation $inv(u, s, t)$. $N$ is the total number of testing samples to be predicted. Both metrics measure the average magnitude of the errors and smaller values indicate better prediction accuracy. Compared to MAE, RMSE gives relatively high weights to large errors and turns to be more suitable when large errors are particularly undesirable. These two metrics have also been adopted by the existing work~\cite{SilicDS13, ZhengL10}.

\subsection{Accuracy Results}\label{sec:accuracy}
We compare \textit{CARP} with the following state-of-the-art approaches that have been recently proposed for reliability prediction of Web services.

\begin{itemize}
\item \textbf{Baseline}: This is a baseline approach that simply uses the overall average value of the observed reliability data as prediction results.
\item \textbf{Hybrid}~\cite{ZhengL10}: This approach models reliability prediction as a collaborative filtering (CF) problem, which is solved by combining two traditional CF approaches: user-based approach (UPCC) and item-based approach (IPCC).
\item \textbf{CLUS}~\cite{SilicDS13}: Based on K-means clustering, this approach clusters historical reliability data according to user-specific, service-specific, and environment-specific parameters respectively, and hashes the average reliability value of each cluster for prediction.
\item \textbf{PMF}~\cite{ZhengL13}: This is a widely-used conventional matrix factorization model, where the reliability data are modelled by a pre-defined low-rank matrix model.
\end{itemize}

As we mentioned before, the observed reliability data are sparse in practice, because each user usually invokes only a small set of services out all of them. To simulate the data sparsity in our experiments, we randomly remove the entries from the data matrix $R$ in our dataset, so that each user only keeps a few available historical reliability records. We use the remaining data for model construction and the removed values for accuracy evaluation. Specifically, we vary the data density from 5\% to 25\% at a step increase of 5\%. Data density = 5\%, for example, indicates that each user invokes only 5\% of the services, and each service is invoked by 5\% of the users. For \textit{Hybrid} and \textit{CLUS}, we employ the executable program with its parameters provided in~\cite{SilicDS13}. For \textit{PMF}, we carefully tune the parameters and set $d = 2$ and $\lambda=0.01$ with best accuracy results. To make \textit{CARP} consistent with other approaches, we set the number of context conditions $C=7$ as with \textit{CLUS}, and set $d = 2$, $\lambda=0.01$ as with \textit{PMF}. Each experiment is run for 20 times and the average results are reported.

Table~\ref{tab:comparison} provides the prediction accuracy results of different approaches in terms of MAE and RMSE. We can see that \textit{CARP} consistently outperforms the other approaches with smaller MAE and RMSE. Compared to the most competitive results of \textit{PMF}, \textit{CARP} still has 13.2\%$\sim$41.0\% improvement on MAE and 6.0\%$\sim$38.9\% improvement on RMSE. It indicates that our \textit{CARP} model fits the reliability data better with context-aware reliability modelling. In particular, CARP achieves a larger improvement at a smaller data density (e.g., the largest improvement is achieved at data density = 5\%), which demonstrates its effectiveness in alleviating the data sparsity problem for reliability prediction. 

\begin{figure}[!t]
  \centering
   \subfloat[MAE]{
  \includegraphics[width=0.32 \textwidth]{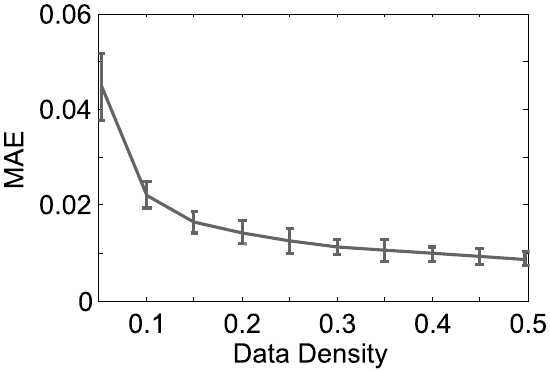}}
  \hfil
   \subfloat[RMSE]{
     \includegraphics[width=0.32 \textwidth]{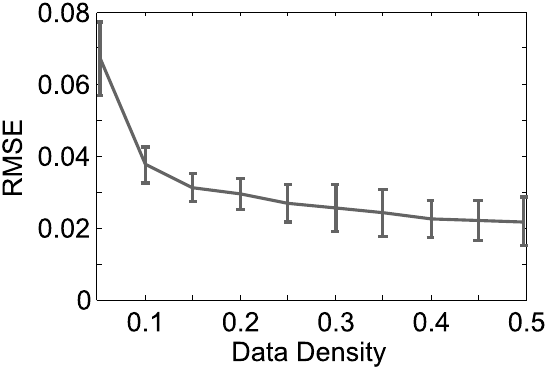}}
  \caption{Impact of Data Sparsity}\label{fig:datadensity}
\end{figure}

\subsection{Impact of Data Sparsity}
To study the impact of data sparsity on prediction accuracy, we evaluate \textit{CARP} by varying the data density from 5\% to 50\% at a step increase of 5\%. A lower data density indicates higher data sparsity because more data are removed during data processing. Fig.~\ref{fig:datadensity} presents the evaluation results (with 95\% confidence interval) on both MAE and RMSE. We can observe that better prediction accuracy can be achieved with the increase of data density from 5\% to 50\%. That is, MAE decreases from 0.045 to 0.009 and RMSE decreases from 0.067 to 0.022. The results show that more training data can usually provide more useful information for model construction and thus achieve better prediction accuracy. In particular, the significant fluctuation of the curve, when the data is extremely sparse (e.g., data density = 5\%), further confirms that data sparsity is a great challenge in achieving accurate reliability prediction. Our \textit{CARP} approach takes a first step forward for addressing the data sparsity challenge and achieves a significant improvement.

\section{Discussion}\label{sec:discussion}

\textbf{Reliability measurement v.s. prediction}.
One may argue that active measurement (e.g., through invoking a service periodically) is a straightforward way to determine reliability of a service. However, this simple approach is not scalable. Each user may have a large number of services to measure, while each service usually has a large user base. Active measurement would incur prohibitive overhead to both users and services. Furthermore, many services are not free, which will lead to additional cost of service invocations. In these cases, it is desired to perform accurate reliability prediction without intensive direct service invocations. 

\textbf{Collaborative data collection}.
A data collection framework capable of assembling invocation records from users is needed to support online reliability prediction. Our work is developed based on underlying usage data collection, but we focus primarily on processing and prediction of reliability values. We have previously developed a WSRec framework~\cite{ZhengMLK11} for collaborative QoS collection, where a set of QoS collector agents were developed using Apache Axis and further deployed on the global PlanetLab platform to collect QoS records of publicly-available Web services at runtime. This framework can be easily extended to support reliability data collection. In addition, a privacy-preserving scheme is explored in~\cite{Zhu_ICWS15_A} by applying differential privacy techniques to dealing with potential privacy issues of usage data collection from different users. We also expect to employ a streaming data platform, e.g., Amazon Kinesis\footnote{\url{https://aws.amazon.com/kinesis}}, to collect real-time usage data streams, but we leave the implementation of such an end-to-end system for future work.

\textbf{Representativeness of datasets}.
CARP is a data-driven approach that highly depends on the characteristics of reliability data. We develop and validate CARP based on the reliability dataset collected from real-world services. But the results may still be limited by the diversity of our dataset. Real-world service usage datasets are scarce in public. To improve generalizability, we plan to further validate our apporach on some other QoS attributes, such as response time and throughput. We believe this work can serve as a good baseline for future research. 


\section{Related Work}\label{sec:relatedwork}

\subsection{Software Quality Assessment}
Software quality assessment~\cite{SunazukaAY85, Kan02} is an important field of study and practice in software engineering that can aid in decision making during all phases of software lifecycle. Software quality~\cite{JungKC04, Gillies11} generally covers many aspects of software products such as correctness, performance, reliability, maintainability, etc. Unfortuntely, many of them cannot be easily quantified or measured for software quality assessment. As a result, an abundance of software quality prediction models~\cite{Kan02} have been built based on measurable internal metrics. Early work applies simple regression models (e.g., multivariate regression model~\cite{AgrestiES90, MunsonK90}) to establishing projections between various design structures and the resulting software quality charactersitics. More recent work (see~\cite{RashidPB12}) proposes the use of sophisticated machine leraning techqniues for improving software quality prediction. These stuides focus on analyzing the expected software quality of design alternatives from the software itself. 

In particualr, as an important aspect of software quality, software reliability assessment and prediction have been extensively studied in the last decades~\cite{Lyu96, Ganesh13}. The researchers have proposed a variety of reliability prediction models, such as Palladio component model~\cite{BroschKBR12}, Poisson process model~\cite{HuangLK03}, structure-based model~\cite{GokhaleL05}, etc. However, most of these existing models target at analyzing traditional white-box software systems, where the reliability of system components are all known. In this paper, we propose to address the problem of user-perceived reliability of black-box services, where existing models are inapplicable. We present a novel approach that can exploit historical usage data from users for context identification of service invocations and can further leverage them for context-aware reliability prediction.


\subsection{QoS Prediction of Web Services}
The work most closely related to ours is the study on QoS (Quality-of-Service) prediction of Web services. Web services are black-box software services that provide software components as building blocks for enterprise application integration (via Web service composition~\cite{ZengBNDKC04}). QoS attributes such as response time, throughput, and reliability are widely used to evaluate the non-functional aspects of Web services for QoS-based Web service composition~\cite{ZengBNDKC04}. To address the QoS prediction problem of Web services, some prior studies have proposed the use of collaborative filtering techniques in recent literature. Collaborative filtering (CF)~\cite{SuK09} is a well-studied technique for rating prediction in recommender systems, which consists of two types of approaches: neighbourhood-based approaches and model-based approaches.  For example, in our previous work we propose neighbourhood-based collaborative filtering approaches (e.g., UIPCC~\cite{ZhengL10, ZhengMLK11}) and model-based collaborative filtering approaches (e.g., PMF~\cite{ZhengL13}, EMF~\cite{XuYWHT14}, AMF~\cite{ZhuHZL14, Zhu_TPDS}) for service quality (or reliability) prediction. However, these models only consider user-specific and service-specific fators, which results in low prediction accuracy. Two more recent stuides~\cite{ZhangZL11, ZhangSLG14} further incorporate temporal information into their models and make use of tensor factorization for time-aware QoS prediction. But tensor factorization suffers from the scalability problem and is not sufficiently efficient for online reliability prediction in our setting. 

Current research has seldom focused on user-perceived reliability prediction of software services. Zheng et al.~\cite{ZhengL10} make the first effort in this direction, where they employ historical usage data from users for reliability prediction and model it as a collaborative filtering problem. In~\cite{ZhengL10}, they propose a neighbourhood-based approach, \textit{Hybrid}, which combines two traditional CF approaches: user-based CF (UPCC) and item-based CF (IPCC). The following work~\cite{ZhengL13} further extends a model-based approach, matrix factorization (PMF), to address this problem. However, these models only consider user-specific and service-specific parameters. Silic et al.~\cite{SilicDS13} make a further step forward and incorporate environment-specific parameters for reliability prediction. This approach achieves scalability by clustering reliability data according to user-specific, service-specific, and environment-specific parameters, but sacrifices prediction accuracy (which is worse than \textit{PMF} as shown in Table~\ref{tab:comparison}). Our approach, instead, addresses these limitations on accuracy and scalability by performing context-aware reliability prediction.

\subsection{Data-Driven Software Engineering}
The data generated throughout the software lifecycle (e.g., source code, revision histories, bug reports, and runtime logs) contain a wealth of valuable information that can aid in software engineering tasks~\cite{ZhengZL13_Ser}. The goal to explore the potential of such rich data motivates a large body of studies related to mining software engineering data\cite{XiePH07}, software intelligence~\cite{HassanX10}, and software analytics~\cite{ZhangX13}. For example, Lessmann et al. study the use of classification models for defect prediction~\cite{LessmannBMP08}. Lin et al. employ clustering techniques to help with system failure diagnosis\cite{LinZLZC16}. Xie et al. employ natural language processing techniuqes to extract method specifications~\cite{PanditaXZXOP12}. Zhou et al. leverage information retrieval techniques for bug localization~\cite{ZhouZL12}. All these studies employ data-driven approaches to gain actionable information and uncover powerful insights for better software development and maintenance. Our work can be viewed as another application in data-driven software engineering, where we describe the novel use of context-specific matrix factorization on historical invocation data for service reliability prediction.

\section{Conclusion and Future Work}\label{sec:conclusion}
This paper presents {CARP}, a context-aware reliability prediction approach for user-perceived reliability prediction of black-box services. {CARP} exploits historical usage data from users to assess the observed reliability of services, and further leverage them to construct context-aware reliability models. Through context-aware model training and prediction, {CARP} is capable of alleviating the data sparsity problem that heavily limits the existing models. The experimental results show that {CARP} makes a significant improvement in prediction accuracy over the state-of-the-art reliability models.

The use of data-driven approaches is promising for the quality management of black-box services in the field. We believe CARP can serve as a good starting point towards this end. As part of our future work, we plan to: \textit{1)} develop more robust reliability prediction approaches to handle the data collection from malicious users and services, \textit{2)} consider data privacy when performing collaborative reliability prediction, and \textit{3)} perform reliability evaluations on real-world services to help identify and address reliability issues.

\section*{Acknowledgment}
The work described in this paper was supported by the National Natural 
Science Foundation of China (Project No. 61332010 and 61502401), the 
National Basic Research Program of China (973 Project No. 2014CB347701), and the Research Grants Council of the Hong Kong Special Administrative Region, China (No. CUHK 14234416 of the General Research
Fund). 

\flushend
{\renewcommand\baselinestretch{1}\setlength{\parsep}{0.5ex}\selectfont
\bibliographystyle{abbrv}
\bibliography{icws}

\begin{thebibliography}{10}

\bibitem{ReliabilityPredictionBasics}
Reliability prediction basics.
\newblock
  \url{http://www.reliabilityeducation.com/ReliabilityPredictionBasics.pdf}.

\bibitem{AgrestiES90}
W.~Agresti, W.~Evanco, and M.~Smith.
\newblock Early experiences building a software quality prediction model.
\newblock In {\em Proc. of the 15th Annual Software Engineering Workshop},
  1990.

\bibitem{BroschKBR12}
F.~Brosch, H.~Koziolek, B.~Buhnova, and R.~Reussner.
\newblock Architecture-based reliability prediction with the palladio component
  model.
\newblock {\em {IEEE} Trans. Software Eng.}, 38(6):1319--1339, 2012.

\bibitem{Gillies11}
A.~Gillies.
\newblock {\em Software quality: theory and management}.
\newblock Lulu.com, 3rd edition, 2011.

\bibitem{GokhaleL05}
S.~S. Gokhale and M.~R. Lyu.
\newblock A simulation approach to structure-based software reliability
  analysis.
\newblock {\em {IEEE} Trans. Software Eng.}, 31(8):643--656, 2005.

\bibitem{HassanX10}
A.~E. Hassan and T.~Xie.
\newblock Software intelligence: the future of mining software engineering
  data.
\newblock In {\em Proc. of the 18th {ACM} {SIGSOFT} International Symposium on
  Foundations of Software Engineering (FSE)}, pages 161--166, 2010.

\bibitem{HuangLK03}
C.~Huang, M.~R. Lyu, and S.~Kuo.
\newblock A unified scheme of some nonhomogenous poisson process models for
  software reliability estimation.
\newblock {\em {IEEE} Trans. Software Eng.}, 29(3):261--269, 2003.

\bibitem{JungKC04}
H.-W. Jung, S.-G. Kim, and C.-S. Chung.
\newblock Measuring software product quality: A survey of {ISO}/{IEC} 9126.
\newblock {\em IEEE Software}, 21(5):88--92, 2004.

\bibitem{Kan02}
S.~H. Kan.
\newblock {\em Metrics and Models in Software Quality Engineering}.
\newblock Addison-Wesley Longman Publishing Co., Inc., 2nd edition, 2002.

\bibitem{LessmannBMP08}
S.~Lessmann, B.~Baesens, C.~Mues, and S.~Pietsch.
\newblock Benchmarking classification models for software defect prediction:
  {A} proposed framework and novel findings.
\newblock {\em {IEEE} Trans. Software Eng.}, 34(4):485--496, 2008.

\bibitem{LinZLZC16}
Q.~Lin, H.~Zhang, J.~Lou, Y.~Zhang, and X.~Chen.
\newblock Log clustering based problem identification for online service
  systems.
\newblock In {\em Proc. of the 38th International Conference on Software
  Engineering (ICSE)}, pages 102--111, 2016.

\bibitem{Lyu96}
M.~R. Lyu, editor.
\newblock {\em Handbook of Software Reliability Engineering}.
\newblock McGraw-Hill, Inc., 1996.

\bibitem{Lyu07}
M.~R. Lyu.
\newblock Software reliability engineering: {A} roadmap.
\newblock In {\em Proc. of ACM/IEEE ICSE, Future of Software Engineering (FOSE)
  Track}, pages 153--170, 2007.

\bibitem{MunsonK90}
J.~Munson and Y.~M. Khoshgoftaar.
\newblock Regression modelling of software quality: Empirical investigation.
\newblock {\em J. Electron. Mater.}, 19(6):106--114, 1990.

\bibitem{Ganesh13}
G.~J. Pai.
\newblock A survey of software reliability models.
\newblock {\em CoRR}, abs/1304.4539, 2013.

\bibitem{PanditaXZXOP12}
R.~Pandita, X.~Xiao, H.~Zhong, T.~Xie, S.~Oney, and A.~M. Paradkar.
\newblock Inferring method specifications from natural language {API}
  descriptions.
\newblock In {\em Proc. of the 34th International Conference on Software
  Engineering ({ICSE})}, pages 815--825, 2012.

\bibitem{RashidPB12}
E.~Rashid, S.~Patnayak, and V.~Bhattacherjee.
\newblock A survey in the area of machine learning and its application for
  software quality prediction.
\newblock {\em {ACM} {SIGSOFT} Software Engineering Notes}, 37(5):1--7, 2012.

\bibitem{SalakhutdinovM07}
R.~Salakhutdinov and A.~Mnih.
\newblock Probabilistic matrix factorization.
\newblock In {\em Proc. of Annual Conference on Neural Information Processing
  Systems (NIPS)}, 2007.

\bibitem{SilicDS13}
M.~Silic, G.~Delac, and S.~Srbljic.
\newblock Prediction of atomic web services reliability based on k-means
  clustering.
\newblock In {\em Proc. of the Joint Meeting of the European Software
  Engineering Conference and the {ACM} {SIGSOFT} Symposium on the Foundations
  of Software Engineering (ESEC/FSE)}, pages 70--80, 2013.

\bibitem{SilicDS15}
M.~Silic, G.~Delac, and S.~Srbljic.
\newblock Prediction of atomic web services reliability for {QoS}-aware
  recommendation.
\newblock {\em {IEEE} Trans. Services Computing}, 8(3):425--438, 2015.

\bibitem{SuK09}
X.~Su and T.~M. Khoshgoftaar.
\newblock A survey of collaborative filtering techniques.
\newblock {\em Adv. Artificial Intellegence}, 2009.

\bibitem{SunazukaAY85}
T.~Sunazuka, M.~Azuma, and N.~Yamagishi.
\newblock Software quality assessment technology.
\newblock In {\em Proc. of the 8th International Conference on Software
  Engineering (ICSE)}, pages 142--149, 1985.

\bibitem{WangLS09}
Y.~Wang, W.~M. Lively, and D.~B. Simmons.
\newblock Web software traffic characteristics and failure prediction model
  selection.
\newblock {\em J. Comp. Methods in Sci. and Eng.}, 9(1,2S1):23--33, 2009.

\bibitem{XiePH07}
T.~Xie, J.~Pei, and A.~E. Hassan.
\newblock Mining software engineering data.
\newblock In {\em Proc. of the 29th International Conference on Software
  Engineering (ICSE)}, pages 172--173, 2007.

\bibitem{XuYWHT14}
Y.~Xu, J.~Yin, Z.~Wu, D.~He, and Y.~Tan.
\newblock Reliability prediction for service oriented system via matrix
  factorization in a collaborative way.
\newblock In {\em Proc. of the 7th {IEEE} International Conference on
  Service-Oriented Computing and Applications ({SOCA})}, pages 125--130, 2014.

\bibitem{ZengBNDKC04}
L.~Zeng, B.~Benatallah, A.~H.~H. Ngu, M.~Dumas, J.~Kalagnanam, and H.~Chang.
\newblock {QoS}-aware middleware for web services composition.
\newblock {\em {IEEE} Trans. Software Eng.}, 30(5):311--327, 2004.

\bibitem{ZhangX13}
D.~Zhang and T.~Xie.
\newblock Software analytics: achievements and challenges.
\newblock In {\em Proc. of the 35th International Conference on Software
  Engineering ({ICSE})}, page 1487, 2013.

\bibitem{ZhangSLG14}
W.~Zhang, H.~Sun, X.~Liu, and X.~Guo.
\newblock Temporal {QoS}-aware web service recommendation via non-negative
  tensor factorization.
\newblock In {\em Proc. of the International World Wide Web Conference
  ({WWW})}, pages 585--596, 2014.

\bibitem{ZhangZL11}
Y.~Zhang, Z.~Zheng, and M.~R. Lyu.
\newblock Wspred: {A} time-aware personalized {QoS} prediction framework for
  web services.
\newblock In {\em Proc. of the 22nd {IEEE} International Symposium on Software
  Reliability Engineering ({ISSRE})}, pages 210--219, 2011.

\bibitem{ZhengL08}
Z.~Zheng and M.~R. Lyu.
\newblock {WS-DREAM:} {A} distributed reliability assessment mechanism for web
  services.
\newblock In {\em Proc. of IEEE/IFIP International Conference on Dependable
  Systems and Networks (DSN)}, pages 392--397, 2008.

\bibitem{ZhengL10}
Z.~Zheng and M.~R. Lyu.
\newblock Collaborative reliability prediction of service-oriented systems.
\newblock In {\em Proc. of the International Conference on Software Engineering
  (ICSE)}, pages 35--44, 2010.

\bibitem{ZhengL13}
Z.~Zheng and M.~R. Lyu.
\newblock Personalized reliability prediction of web services.
\newblock {\em {ACM} Trans. Softw. Eng. Methodol.}, 22(2):12, 2013.

\bibitem{ZhengMLK11}
Z.~Zheng, H.~Ma, M.~R. Lyu, and I.~King.
\newblock {QoS}-aware web service recommendation by collaborative filtering.
\newblock {\em {IEEE} T. Services Computing}, 4(2):140--152, 2011.

\bibitem{ZhengZL13_Ser}
Z.~Zheng, J.~Zhu, and M.~R. Lyu.
\newblock Service-generated big data and big data-as-a-service: An overview.
\newblock In {\em Proc. of the {IEEE} International Congress on Big Data
  (BigData Congress)}, pages 403--410, 2013.

\bibitem{ZhouZL12}
J.~Zhou, H.~Zhang, and D.~Lo.
\newblock Where should the bugs be fixed? more accurate information
  retrieval-based bug localization based on bug reports.
\newblock In {\em Proc. of the 34th International Conference on Software
  Engineering ({ICSE})}, pages 14--24, 2012.

\bibitem{wsdream_repo}
J.~Zhu, P.~He, Z.~Zheng, and M.~R. Lyu.
\newblock {WS-DREAM}: Towards open datasets and source code for web service
  recommendation. {Available} at \url{http://wsdream.github.io}.

\bibitem{ZhuHZL14}
J.~Zhu, P.~He, Z.~Zheng, and M.~R. Lyu.
\newblock Towards online, accurate, and scalable {QoS} prediction for runtime
  service adaptation.
\newblock In {\em Proc. of the 34th {IEEE} International Conference on
  Distributed Computing Systems ({ICDCS})}, pages 318--327, 2014.

\bibitem{Zhu_ICWS15_A}
J.~Zhu, P.~He, Z.~Zheng, and M.~R. Lyu.
\newblock A privacy-preserving {QoS} prediction framework for web service
  recommendation.
\newblock In {\em Proc. of the {IEEE} International Conference on Web Services
  ({ICWS})}, pages 241--248, 2015.

\bibitem{Zhu_TPDS}
J.~Zhu, P.~He, Z.~Zheng, and M.~R. Lyu.
\newblock Online {QoS} prediction for runtime service adaptation via adaptive
  matrix factorization.
\newblock {\em IEEE Trans. Parallel Distrib. Syst.}, preprint, 2017.

\bibitem{ZhuZZL13_Scale}
J.~Zhu, Z.~Zheng, Y.~Zhou, and M.~R. Lyu.
\newblock Scaling service-oriented applications into geo-distributed clouds.
\newblock In {\em Proc. of the 7th {IEEE} International Symposium on
  Service-Oriented System Engineering}, pages 335--340, 2013.

\end{thebibliography}
\par}

\end{document}